\documentclass[reprint,amsmath,amssymb,aps,pra]{revtex4-1}

\usepackage{graphicx}
\usepackage{bm}
\usepackage{color, soul}

\usepackage{wasysym}
\usepackage{subfigure}
\usepackage{commath}
\usepackage{braket}
\usepackage{amsmath}
\usepackage{amsfonts}
\usepackage{amssymb}
\usepackage{amsthm}

\usepackage{siunitx}
\usepackage{multirow}
\usepackage{epstopdf}
\usepackage{array}
\usepackage{latexsym}
\usepackage[latin1]{inputenc}
\usepackage{textcomp}
\usepackage{ulem}
\usepackage{braket}
\usepackage{siunitx}
\DeclareSIUnit\dBm{dBm}
\usepackage{booktabs}

\begin{document}

\title{High-performance coherent population trapping clock with polarization modulation}

\author{Peter Yun$^1$}
\thanks{E-mail: enxue.yun@obspm.fr, permanent e-mail: yunenxue@163.com}
\author{Fran\c{c}ois Tricot$^1$}

\author{Claudio Eligio Calosso$^2$}
\author{Salvatore Micalizio$^2$}
\author{Bruno Fran\c{c}ois$^3$}
\author{Rodolphe Boudot$^3$}
\author{St\'ephane Gu\'erandel$^1$}
\author{Emeric de Clercq$^1$}

\address{$^1$ LNE-SYRTE, Observatoire de Paris, PSL Research University, CNRS, Sorbonne Universit\'es, UPMC Univ. Paris 06, 61 avenue de l'Observatoire, 75014 Paris, France}
\address{$^2$ Istituto Nazionale di Ricerca Metrologica, INRIM, Strada delle Cacce 91, 10135 Torino, Italy}
\address{$^3$ FEMTO-ST, CNRS, UFC, 26 chemin de l'\'Epitaphe 25030 Besan\c{c}on Cedex, France}

\date{\today}%

\begin{abstract}

We demonstrate a vapor cell atomic clock prototype based on continuous-wave (CW) interrogation and double-modulation coherent population trapping (DM-CPT) technique. The DM-CPT technique uses a synchronous modulation of polarization and relative phase of a bi-chromatic laser beam in order
to increase the number of atoms trapped in a dark state, \textit{i.e.} a non-absorbing state. The narrow resonance, observed in transmission of a Cs vapor cell, is used as a narrow frequency discriminator in an atomic clock. A detailed characterization of the CPT resonance versus numerous parameters is reported. A short-term frequency stability of $3.2 \times 10^{-13} \tau^{-1/2}$ up to \SI{100}{\second} averaging time is measured. 
These performances are more than one order of magnitude better than industrial Rb clocks and comparable to those of best laboratory-prototype vapor cell clocks. The noise budget analysis shows that the short and mid-term frequency stability is mainly limited by the power fluctuations of the microwave used to generate the bi-chromatic laser. These preliminary results demonstrate that the DM-CPT technique is well-suited for the development of a high-performance atomic clock, with potential compact and robust setup due to its linear architecture. This clock could find future applications in industry, telecommunications, instrumentation or global navigation satellite systems.
\end{abstract}
\maketitle

\section{Introduction}
\par Microwave Rb vapor-cell atomic clocks \cite{Camparo07}, based on optical-microwave double resonance, 
are today ubiquitous timing devices used in numerous fields of industry including instrumentation, 
telecommunications or satellite-based navigation systems. 
Their success is explained by their ability to demonstrate excellent short-term fractional frequency stability at the level of 10$^{-11} \tau^{-1/2}$, combined with a small size, weight, power consumption and a relatively modest cost. 
Over the last decade, the demonstration of advanced atom interrogation techniques (including for instance pulsed-optical-pumping (POP)) 
using narrow-linewidth semiconductor lasers has conducted to the development in laboratory of new-generation vapor cell clocks \cite{Micalizio12, Bandi14, Kang15, Micalizio:RN2016}. 
These clocks have succeeded to achieve a 100 times improvement in frequency stability compared to existing commercial vapor cell clocks.
\par In this domain, clocks based on a different phenomenon, named coherent population trapping (CPT), 
have proven to be promising alternative candidates. Since its discovery in $1976$ \cite{Alzetta76}, 
coherent population trapping physics \cite{Arimondo96, Bergmann98, Wynands99, Vanier05} has motivated stimulating studies in various fields covering fundamental and applied physics such as slow-light experiments \cite{Bajcsy}, 
high-resolution laser spectroscopy, magnetometers \cite{Schwindt, Breschi}, 
laser cooling \cite{Aspect} or atomic frequency standards. 
Basically, CPT occurs by connecting two long-lived ground state hyperfine levels of an atomic specie to a common excited state by simultaneous action of two resonant optical fields. At null Raman detuning, i.e. when the frequency difference between both optical fields matches perfectly the atomic ground-state hyperfine frequency, atoms are trapped through a destructive quantum interference process into a noninteracting coherent superposition of both ground states, so-called dark state, resulting in a clear decrease of the light absorption or equivalently in a net increase of the transmitted light. The output resonance signal, whose line-width is ultimately limited by the CPT coherence lifetime, can then be used as a narrow frequency discriminator towards the development of an atomic frequency standard. In a CPT-based clock, unlike the traditional double-resonance Rb clock \cite{Vanier07}, the microwave signal used to probe the hyperfine frequency is directly optically carried allowing to remove the microwave cavity and potentially to shrink significantly the clock dimensions.
\par The application of CPT to atomic clocks was firstly demonstrated in a sodium atomic beam \cite{Thomas81,Thomas82}. 
In $1993$, N. Cyr et al proposed a simple method to produce a microwave clock transition in a vapor cell with purely optical means by using a modulated diode laser \cite{Cyr93}, demonstrating its high-potential for compactness.  
In $2001$, a first remarkably compact atomic clock prototype was demonstrated in NIST \cite{Kitching00,Kitching01}. Further integration was achieved later thanks to the proposal \cite{Kitching02} and development of micro-fabricated alkali vapor cells \cite{Liew04}, 
leading to the demonstration of the first chip-scale atomic clock prototype (CSAC) \cite{Knappe04} and later to the first commercially-available CSAC \cite{csac11}. 
Nevertheless, this extreme miniaturization effort induces a typical fractional frequency stability limited at the level of $10^{-10} \tau^{-1/2}$, not compliant with dedicated domains requiring better stability performances. In that sense, in the frame of the European collaborative MClocks project \cite{Mclocks}, significant efforts have been pursued to demonstrate compact high-performance CPT-based atomic clocks and to help to push this technology to industry. 
\par In standard CPT clocks, a major limitation to reach better frequency stability performances is the low contrast ($C$, the amplitude-to-background ratio) of the detected CPT resonance. This low contrast is explained by the fact that atoms interact with a circularly polarized bichromatic laser beam, leading most of the atomic population into extreme Zeeman sub-levels of the ground state, so called "end-states". 
Several optimized CPT pumping schemes, aiming to maximize the number of atoms participating to the clock transition, 
have been proposed in the literature to circumvent this issue (\cite{Shah10,Yun12}, and references therein), 
but at the expense of increased complexity.
\par In that sense, a novel constructive polarization modulation CPT \cite{Yun14} pumping technique, named double-modulation (DM) scheme, was recently proposed. It consists to apply a phase modulation between both optical components of the bichromatic laser synchronously with a polarization modulation. The phase modulation is needed to ensure a common dark state to both polarizations, allowing
 to pump a maximum number of atoms into the desired magnetic-field insensitive clock state. This elegant solution presents the main advantage compared to the push-pull optical pumping \cite{Jau04,Liu13,Hafiz15} or the lin$\perp$lin technique \cite{Zanon05,Danet14} to avoid any optical beam separation or superposition and is consequently well-adapted to provide a compact and robust linear architecture setup.
\par In this article, we demonstrate a high-performance CW-regime CPT clock based on the DM technique. 
Optimization of the short-term frequency stability is performed by careful characterization of the CPT resonance versus 
relevant experimental parameters. 
A short-term frequency stability at the level of $3.2 \times 10^{-13} \tau^{-1/2}$ up to 100 s, comparable to best vapor cell frequency standards, is reported. A detailed noise budget is given, highlighting a dominant contribution of the microwave power fluctuations. Section II describes the experimental setup. Section III reports the detailed CPT resonance spectroscopy versus experimental parameters. Section IV reports best short-term frequency stability results. Noise sources limiting the stability are carefully analysed. In section V, we study the clock frequency shift versus each parameter and estimate the limitation of the clock mid-term frequency stability.

\section{Experimental set-up}

\subsection{Optical set-up}

\par
\begin{figure}[h]
\centering
\includegraphics[width=0.45\textwidth]{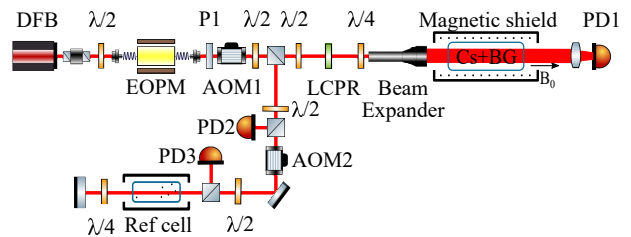}
\caption{(Color online) Optical setup for DM CPT, where the abbreviations stand for:
DFB distributed feedback diode laser, EOPM electro-optic phase modulator, LCPR liquid crystal polarization rotator, 
AOM acousto-optic modulator, P polarizer, $\lambda/2(4)$ half(quarter)-wave plate, BG buffer gas, PD photodiode.}
\label{fig1}
\end{figure}

 Our setup is depicted in Fig.~\ref{fig1}. A DFB laser diode emits a monochromatic laser beam around \SI{895}{\nano\meter},
the wavelength of the Cs $D_1$ line. With the help of a fiber electro-optic phase modulator (EOPM),
modulated at \SI{4.6}{\giga\hertz} with about \SI{26}{\dBm} microwave power,
about $70\%$ of the carrier power is transferred into both first-order sidebands used for CPT interaction.
The phase between both optical sidebands, so-called Raman phase in the following,
is further modulated through the driving \SI{4.6}{\giga\hertz} microwave signal.
Two acousto-optic modulators (AOMs) are employed. The first one, AOM1, is used for laser power stabilization. 
The second one, AOM2, allows to compensate for the buffer-gas induced optical frequency shift ($\approx$ \SI{160}{\mega\hertz}) in the CPT clock cell.
A double-modulated laser beam is obtained by combining the phase modulation with a synchronized polarization modulation performed thanks to a liquid crystal polarization rotator (LCPR).
The laser beam is expanded to \SI{9x16}{\milli\meter} before the vapor cell.
The cylindrical Cs vapor cell, \SI{25}{\milli\meter} diameter and \SI{50}{\milli\meter} long,
is filled with \SI{15}{Torr} of mixed buffer gases (argon and nitrogen).
Unless otherwise specified, the cell temperature is stabilized to about \SI{35}{\degreeCelsius}. A uniform magnetic field of \SI{3.43}{\micro\tesla} is applied along the direction of the cell axis by means of a solenoid. The ensemble is surrounded by two magnetic shields in order to remove the Zeeman degeneracy.

\subsection{Fiber EOPM sidebands generation}

\par
\begin{figure}[htb]
\centering
\includegraphics[width=0.45\textwidth]{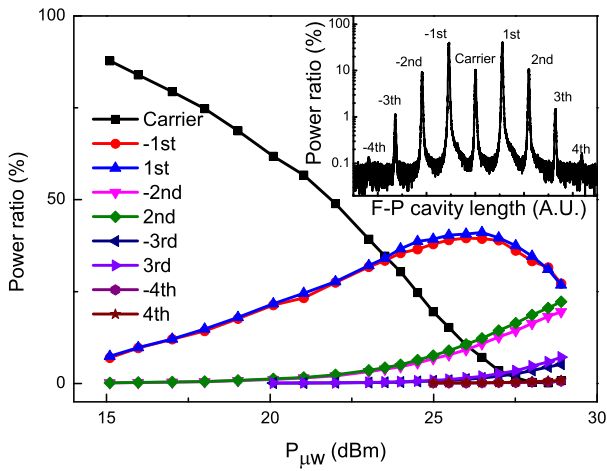}
\caption{(Color online) Fractional power of laser sidebands at the EOPM output 
as a function of \SI{4.596}{\giga\hertz} microwave power. 
Inset, the laser sidebands spectrum with $P_{\mu w}= \SI{26.12}{\dBm}$ obtained by scanning the FP cavity length, 
notice the log scale of the y-axis. }
\label{sideband}
\end{figure}

 We first utilized a Fabry-Perot cavity to investigate the EOPM sidebands power ratio
versus the coupling \SI{4.596}{\giga\hertz} microwave power ($P_{\mu w}$), see Fig.~\ref{sideband}.
We choose $P_{\mu w}$ around \SI{26}{\dBm} to maximize the power transfer efficiency into the first-order sidebands.
The sidebands spectrum is depicted in the inset of Fig.~\ref{sideband}. 

\subsection{Laser power locking}

\begin{figure}[htb]
\centering
\includegraphics[width=0.45\textwidth]{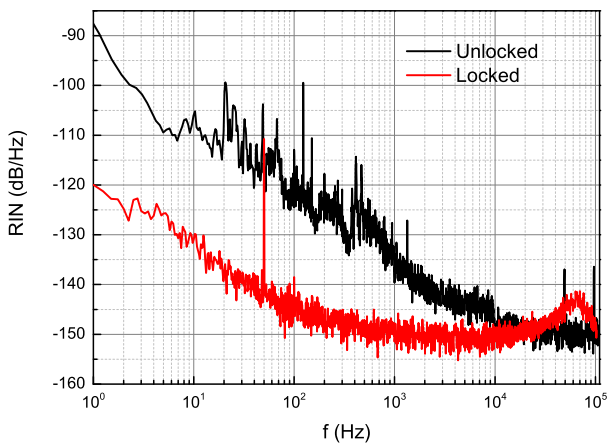}
\caption{(Color online)Power spectral density of the laser RIN with and w/o power locking.}
\label{fig3}
\end{figure}

\par Since the laser intensity noise is known as being one of the main noise sources which limit the performances of a CPT clock \cite{Vanier05,Hafiz15}, laser power needs to be carefully stabilized. For this purpose, a polarization beam splitter (PBS) reflects towards a photo-detector a part of the laser beam, the first-order diffracted by AOM1 following the EOPM. The output voltage signal is compared to an ultra-stable voltage reference (LT1021). The correction signal is applied on a voltage variable power attenuator
set on the feeding RF power line of the AOM1 with a servo bandwidth of about \SI{70}{\kilo\hertz}. 
The out-loop laser intensity noise (RIN) is measured just after the first PBS with a photo-detector (PD), 
which is not shown in Fig.~\ref{fig1}. 
The spectrum of the resulting RIN with and w/o locking is shown in Fig.~\ref{fig3}. 
A \SI{20}{\decibel} improvement at $F_M = \SI{125}{\hertz}$ (LO modulation frequency for clock operation) is obtained in the stabilized regime.

\par It is worth to note that the DFB laser diode we used, with a linewidth of about \SI{2}{\mega\hertz},
is sensitive even to the lowest levels of back-reflections \cite{Schmeissner16}, e.g., the coated collimated lens may introduce some intensity and frequency noise at the regime of \SIrange{0.1}{10}{\kilo\hertz}. Finding the correct lens alignment to minimize the reflection induced noise while keeping a well-collimated laser beam was not an easy task. To reduce light feedback from the EOPM fiber face, we use a \SI{60}{\decibel} isolator before the EOPM.
The fiber-coupled EOPM induces additional intensity noise depicted in Fig.~\ref{fig3}. 
Nevertheless, thanks to the laser power locking, 
we can reduce most of these noises by at least \SI{15}{\decibel} in the range of \SI{1}{\hertz} to \SI{1}{\kilo\hertz}.

\subsection{Laser frequency stabilization}
\begin{figure}[htb]
\centering
\includegraphics[width=0.45\textwidth]{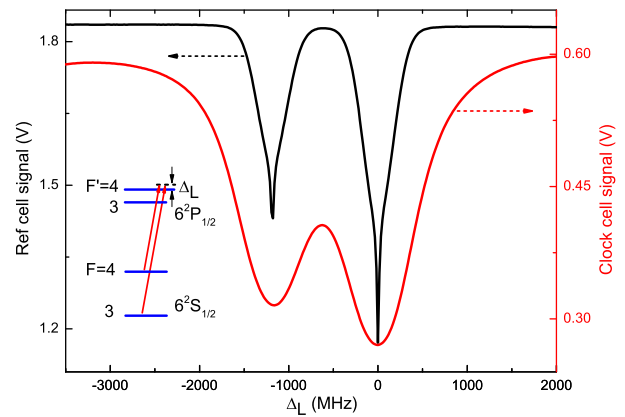}
\caption{(Color online) Spectrum of the Cs $D_1$ line in the vacuum reference cell and in the clock cell recorded with the bichromatic laser.
The two absorptions from left to right correspond to the excited level $\ket{6^2P_{1/2}, F'=3}$ and $\ket{6^2P_{1/2}, F'=4}$, respectively.
For the reference cell signal: laser power \SI{0.74}{\milli\watt}, beam diameter \SI{2}{\milli\meter}, 
cell temperature \SI{22}{\degreeCelsius}, AOM frequency \SI{160}{\mega\hertz}. Inset, the atomic levels involved in $D_1$ line of Cesium. }
\label{fig4}
\end{figure}

\begin{figure}[htb]
\centering
\includegraphics[width=0.45\textwidth]{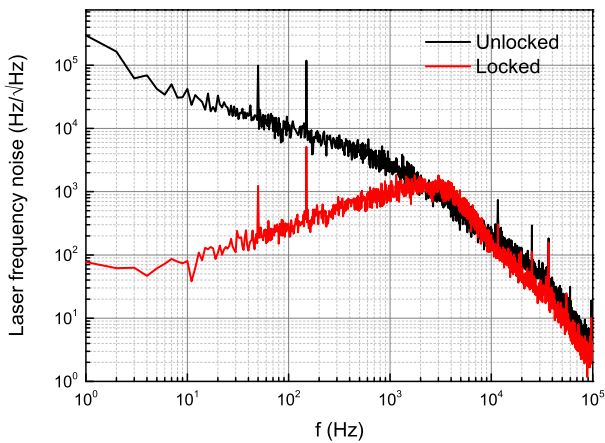}
\caption{(Color online) The laser frequency noise measured on the error signal of the Doppler-free spectrum with and w/o frequency locking.}
\label{fig5}
\end{figure}

Our laser frequency stabilization setup, similar to \cite{Hafiz15,Hafiz16}, is depicted in Fig.~\ref{fig1}. We observe in a vacuum cesium cell the two-color Doppler-free spectrum  depicted in Fig.~\ref{fig4}. 
The bi-chromatic beam, linearly polarized, is retro-reflected after crossing the cell with the orthogonal polarization. 
Only atoms of null axial velocity are resonant with both beams. 
Consequently, CPT states built by a beam are destroyed by the reversed beam, 
leading to a Doppler-free enhancement of the absorption \cite{Hafiz16}.

\par The laser frequency detuning $\Delta_L=0$ in Fig.~\ref{fig4} corresponds to the laser carrier frequency tuned to the center of both transitions
$\ket{6^2S_{1/2}, F=3}\rightarrow\ket{6^2P_{1/2}, F'=4}$  and  $\ket{6^2S_{1/2}, F=4}\rightarrow\ket{6^2P_{1/2}, F'=4}$ in D$_1$ line of Cesium,
where $F$ is the hyperfine quantum number. For this record, the microwave frequency is \SI{4.596}{\giga\hertz}, half the Cs ground state splitting, and the DFB laser frequency is scanned. The frequency noise with and w/o locking are presented in Fig.~\ref{fig5}. 
The servo bandwidth is about \SI{3}{\kilo\hertz} and the noise is found to be reduced by about \SI{25}{\decibel} at $f$ = \SI{125}{\hertz} (local oscillator modulation frequency in clock operation).

\subsection{Polarization modulation}
\begin{figure}[htb]
\centering
\includegraphics[width=0.45\textwidth]{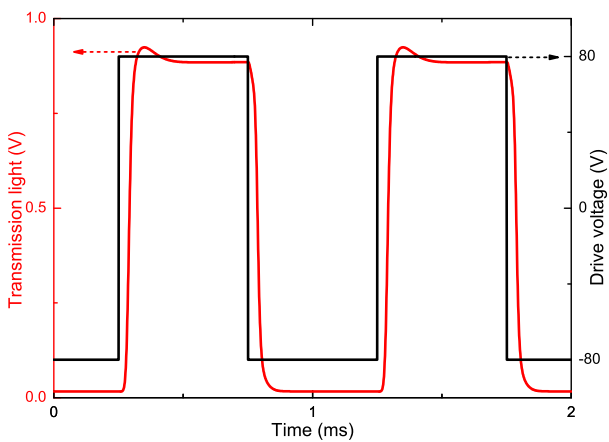}
\caption{(Color online) LCPR response. }
\label{fig6}
\end{figure}

\par We studied the response time of the LCPR (FPR-100-895, Meadowlark Optics). 
As illustrated in Fig.~\ref{fig6}, the measured rise (fall) time is about \SI{100}{\micro\second} and the polarization extinction ratio is about $50$. 
In comparison, the electro-optic amplitude modulator (EOAM) used as polarization modulator in our previous investigations \cite{Yun14} showed a response time of \SI{2.5}{\micro\second} (limited by our high voltage amplifier) and a polarization extinction ratio of $63$. 
Here, we replace it by a liquid crystal device because its low voltage and small size  would be an ideal choice for a compact CPT clock, 
and we will show in the following that the longer switching time does not limit the contrast of the CPT signal.

\subsection{Microwave source and clock servo-loop }

\begin{figure}[htb]
\centering
\includegraphics[width=0.45\textwidth]{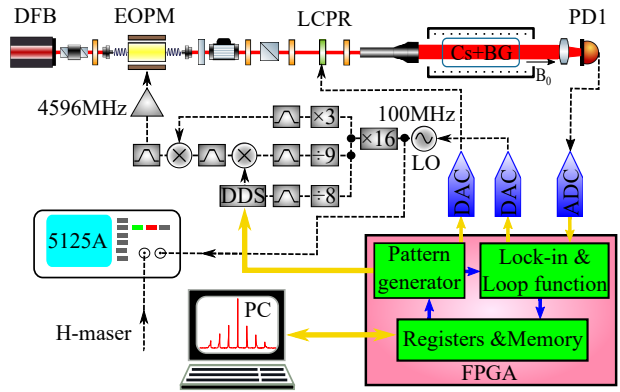}
\caption{(Color online) Electronic architecture of the DM CPT clock: 
DDS direct digital synthesizer, 
DAC digital-to-analog converter, 
ADC analog-to-digital converter, 
FPGA field programmable gate array, LO local oscillator. }
\label{Microwave-chain}
\end{figure}

\par The electronic system (local oscillator and digital electronics for clock operation) used in our experiment is depicted in 
Fig.~\ref{Microwave-chain}. The \SI{4.596}{\giga\hertz} microwave source is based on the design described in \cite{Francois15}.
The local oscillator (LO) is a module (XM16 Pascall) integrating an ultra-low phase noise 100 MHz quartz oscillator
frequency-multiplied without excess noise to \SI{1.6}{\giga\hertz}. 
The \SI{4.596}{\giga\hertz} signal is synthesized by a few frequency multiplication, division and mixing stages.
The frequency modulation and tuning is yielded by a direct digital synthesizer (DDS) referenced to the LO.
The clock operation \cite{Micalizio12,Calosso07} is performed by a single
field programmable gate array (FPGA) which coordinates the operation of the
DDS, analog-to-digital converters (ADC) and digital-to-analog converter (DACs):
\par(1) the DDS generates a signal with phase modulation (modulation rate $f_m$, depth $\pi/2$) and frequency modulation ($F_M$,  depth $\Delta_{F_M}$).
\par(2) the DAC generates a square-wave signal to drive the LCPR with the same rate $f_m$, synchronous to the phase modulation.
\par(3) the ADC is the front-end of the lock-in amplifier.
Another DAC, used to provide the feedback to the local oscillator frequency, is also implemented in the FPGA.

\par The clock frequency is measured by comparing the LO signal with a \SI{100}{\mega\hertz} signal delivered by a H maser of the laboratory in a Symmetricom $5125$A Allan deviation test set. The frequency stability of the maser is $1\times 10^{-13}$ at \SI{1}{\second} integration time.

\section{Clock signal optimization}
\subsection{Time sequence and figure of merit}

\begin{figure}[htb]
\centering
\includegraphics[width=0.45\textwidth]{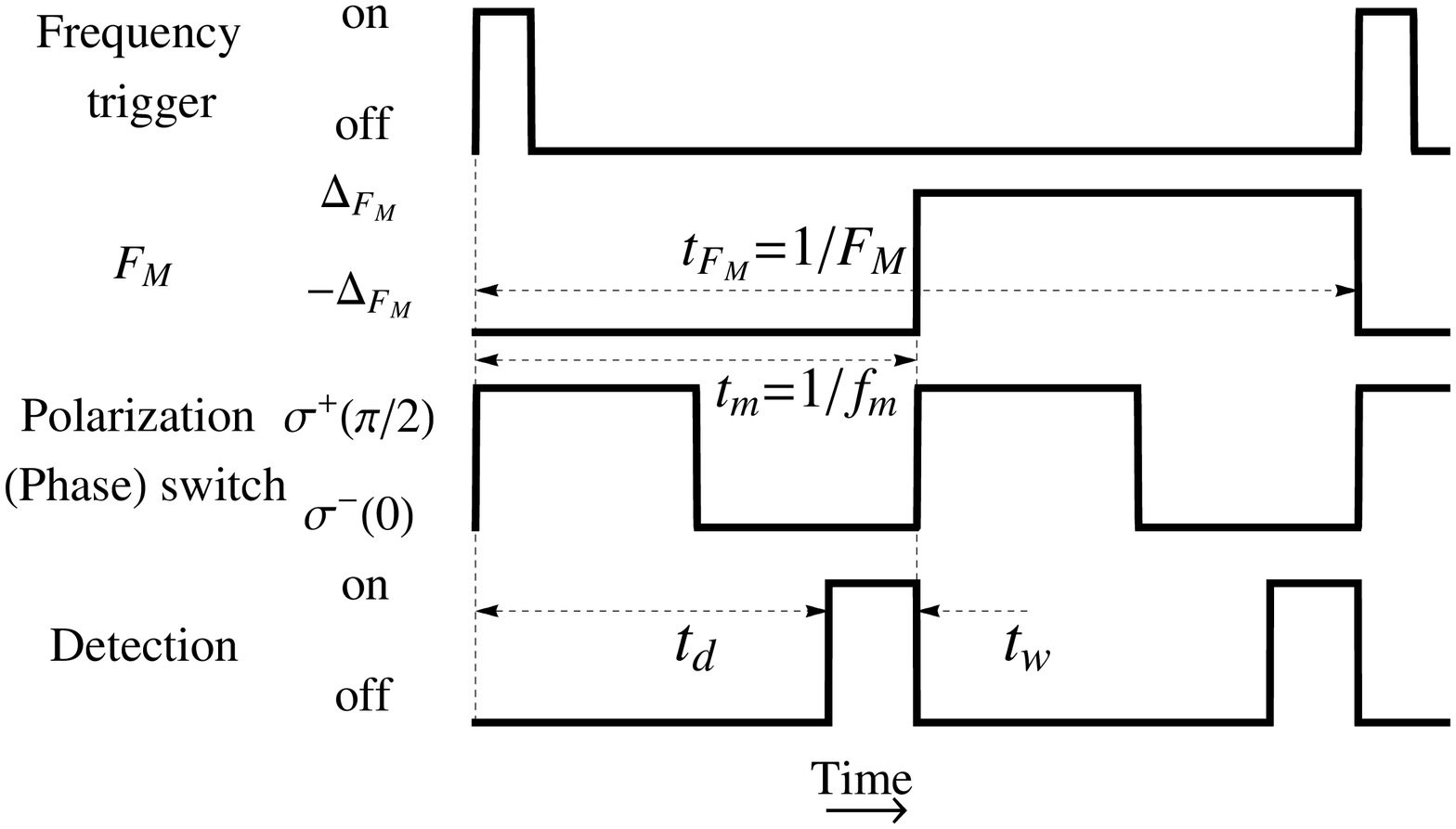}
\caption{ Time sequence. $F_M$ modulation frequency of the \SI{4.596}{\giga\hertz} signal, 
$f_m$ polarization and phase modulation frequency, $t_d$ pumping time, $t_w$ detection window.}
\label{fig8}
\end{figure}

\par As illustrated in Fig.~\ref{fig8}, the polarization and phase modulation share the same modulation function. 
After a pumping time $t_d$ to prepare the atoms into the CPT state, we detect the CPT signal with a window of length $t_w$. 
In order to get an error signal to close the clock frequency loop, the microwave frequency is square-wave modulated with a frequency $F_M$, 
and a depth $\Delta_{F_M}$. 
In our case, we choose $F_M=125$ Hz, as a trade-off between a low frequency to have time to accumulate the atomic 
population into the clock states by the DM 
scheme and a high operating frequency to avoid low frequency noise in the lock-in amplification process 
and diminish the intermodulation effects.

\begin{figure}[h]
\centering
\includegraphics[width=0.45\textwidth]{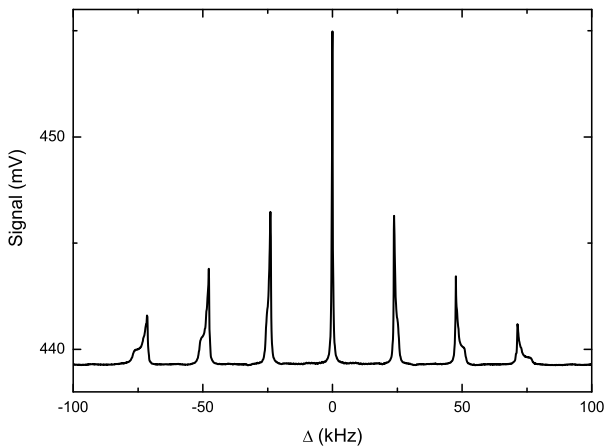}
\caption{ Zeeman spectrum. The center peak is the $0-0$ clock transition. 
Working parameters: 
$t_d= \SI{2}{\milli\second}$, $t_w= \SI{2}{\milli\second}$, $f_m= \SI{500}{\hertz}$, 
$P_L = \SI{163}{\micro\watt}$, $P_{\mu w}= \SI{26.12}{\dBm}$, $T_{cell}=\SI{35.1}{\degreeCelsius}$.  }
\label{Zeeman}
\end{figure}

\par A typical experimental CPT signal, recorded with this time sequence, 
showing all the CPT transitions allowed between Zeeman sub-levels of the Cs ground state is reported in Fig.~\ref{Zeeman}. 
The Raman detuning $\Delta$ is the difference between the two first sideband spacing and the Cs clock resonance. 
The spectrum shows that the clock levels ($0-0$ transition) are the most populated and that the atomic population is symmetrically distributed around the $(m=0)$ sub-levels. The distortion of neighbouring lines is explained by magnetic field inhomogeneities.

\par It can be shown that the clock short-term frequency stability limited by an amplitude noise scales as 
$W_h/C$ \cite{Vanier05}, with $W_h$ the full width at half maximum (FWHM) of the clock resonance, and $C$ the contrast of the resonance. 
Usually, the ratio of contrast $C$ to $W_h$  
is adopted as a figure of merit, \textit{i.e.} $F_C=C/W_h$. The best stability should be obtained by maximizing $F_C$. 

\begin{figure}[h]
\centering
\includegraphics[width=0.45\textwidth]{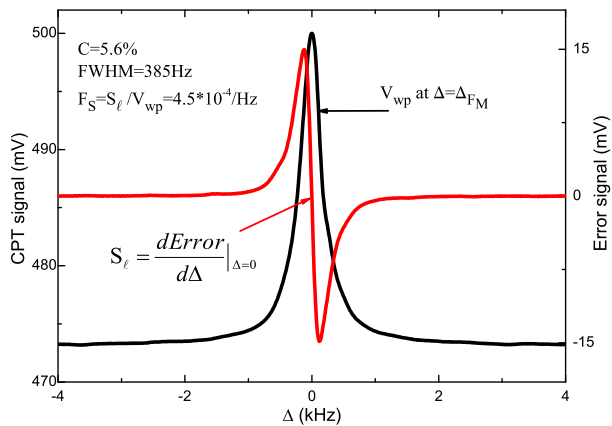}
\caption{(Color online) Signal of the clock transition and error signal. 
Working parameters: $t_d= \SI{3}{\milli\second}$, $t_w= \SI{1}{\milli\second}$, $f_m= \SI{250}{\hertz}$,
$P_L = \SI{163}{\micro\watt}$, $P_{\mu w}= \SI{26.12}{\dBm}$, $T_{cell}=\SI{35.1}{\degreeCelsius}$.}
\label{CPTsignal}
\end{figure}

The stability of the clock is measured by the Allan standard deviation $\sigma_y(\tau)$, 
with $\tau$ the averaging time. 
When the signal noise is white, with standard deviation $\sigma_N$, and for a square-wave frequency modulation, 
the stability limited by the signal-to-noise ratio is equal to \cite{Vanier89} 
\begin{equation}
\centering
\sigma_y(\tau)= \frac{1}{f_c}\frac{\sigma_N}{S_\ell}\sqrt{\frac{1}{\tau}},
\label{sigma_y}
\end{equation}
with $f_c$ the clock frequency, and $S_\ell$ the slope of the frequency discriminator. 
In CPT clocks, one of the main sources of noise is the laser intensity noise, 
which leads to a signal noise proportional to the signal. 
Therefore it is more convenient to characterize the quality of the signal of a CPT atomic clock by a new figure of merit, $F_S=S_\ell/V_{wp}$, where $S_\ell$ is the slope of the error signal (in V/Hz) at Raman resonance ($\Delta=0$), and $V_{wp}$ 
is the detected signal value (in V) at the interrogating frequency (the clock resonance frequency plus the modulation depth $\Delta_{F_M}$), see Fig.~\ref{CPTsignal}. 
Note that an estimation of the discriminator slope is also included in $F_C$, since the contrast is the signal amplitude $A$ divided by the background $B$. $F_C$ then equals $(A/W_h)/B$,  $(A/W_h)$ is a rough approximation of the slope $S_\ell$, and $B$ an approximation of the working signal $V_{wp}$. In our experimental conditions $A/W_h \sim S_\ell/3$.

\par We investigated the effect of relevant parameters on both figures of merit to optimize the clock performances. 
In order to allow a comparison despite different conditions, the error signals are generated with the same unit gain. 
Since the resonance linewidth is also subject to change, it is necessary to optimize the 4.6 GHz 
modulation depth $\Delta_{F_M}$ to maximize $F_S$. Here for simplicity, 
we first recorded the CPT signal, then we can numerically compute optimized values of $\Delta_{F_M}$ and  $F_S$. 
\par In the following, we investigate the dependence of $F_C$ and $F_S$ on several parameters including 
the cell temperature ($T_{cell}$), the laser power ($P_L$), the microwave power ($P_{\mu w}$), 
the detection window duration ($t_w$), the detection start time ($t_d$),
and the polarization and phase modulation frequency ($f_m$).
 
\subsection{Cell temperature and laser power }

\begin{figure}[h]
\centering
\includegraphics[width=0.45\textwidth]{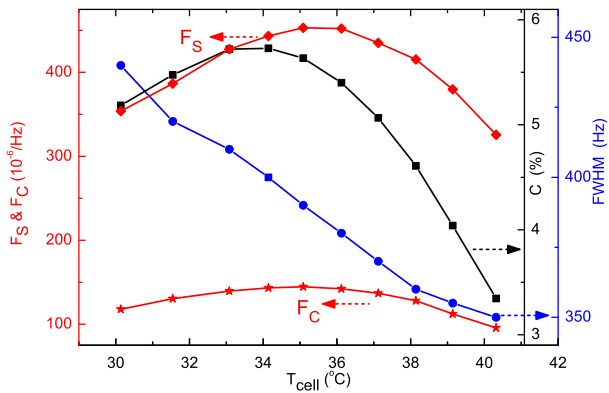}
\caption{(Color online)  Figures of merit $F_S$, $F_C$, contrast $C$ and width of the clock transition 
as function of cell temperature $T_{cell}$. All other working parameters is the same as Fig.~10.}
\label{Tcell}
\end{figure}

\par
\begin{figure}[h]
\centering
\includegraphics[width=0.45\textwidth]{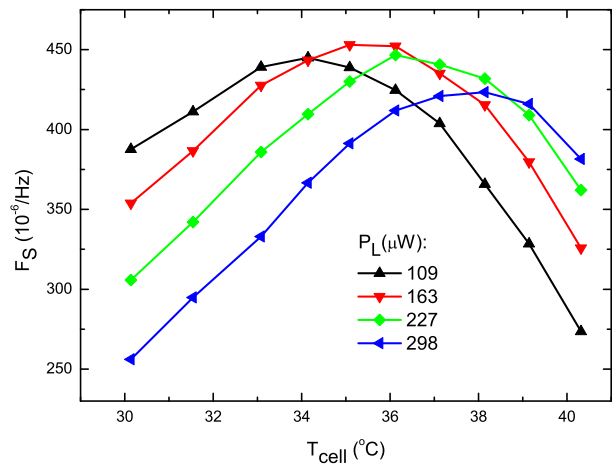}
\caption{(Color online) $F_S$ as a function of cell temperature $T_{cell}$ for various laser powers. 
All other working parameters is the same as Fig.~10. }
\label{Tcell-2}
\end{figure}

From the figures of merit shown in Fig.~\ref{Tcell}, the optimized cell temperature is 
around $T_{cell}=$ \SI{35.1}{\degreeCelsius} for $P_L = \SI{163}{\micro\watt}$.  
The narrower linewidth observed at higher $T_{cell}$, already observed  
by Godone \textit{et al.} \cite{Godone02}, can be explained by the propagation effect: 
the higher the cell temperature, the stronger the light absorption by more atoms, 
and less light intensity is seen by the atoms at the end side of the vapour cell. 
This leads to a reduction of the power broadening and a narrower signal, measured by the transmitted light amplitude. 
The optimum temperature depends on the laser power as depicted in Fig.~\ref{Tcell-2}. 
Nevertheless, the overall maximum of $F_S$ is reached with $P_L = \SI{163}{\micro\watt}$ 
at $T_{cell}=\SI{35.1}{\degreeCelsius}$.

\par
\begin{figure}[hb]
\centering
\includegraphics[width=0.45\textwidth]{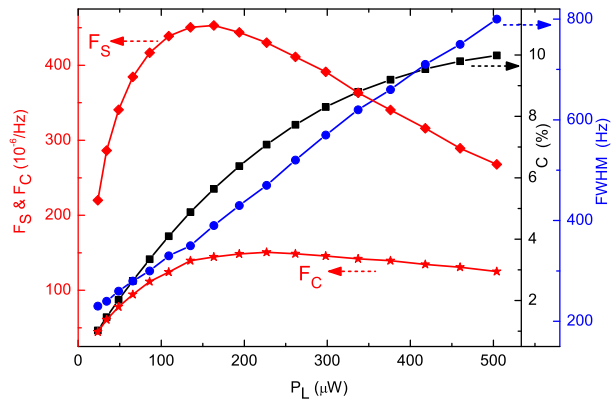}
\caption{(Color online)  $F_S$, $F_C$, $C$ and width of the clock transition as a function of laser power $P_L$ with 
$T_{cell}=\SI{35.1}{\degreeCelsius}$. All other working parameters is the same as Fig.~10.}
\label{PL}
\end{figure}

  The figures of merit $F_S$, $F_C$, the contrast $C$ and the width are plotted as a function of the laser power in Fig.~\ref{PL} for  $T_{cell}=\SI{35.1}{\degreeCelsius}$. The laser powers maximizing  $F_S$ and $F_C$ are $P_L = \SI{163}{\micro\watt}$ and $P_L = \SI{227}{\micro\watt}$, respectively. The Allan deviation reaches a better value at $P_L = \SI{163}{\micro\watt}$, which justifies our choice of  $F_S$ as figure of merit. 
And in the following parameters investigation, we only show the $F_S$ as figure of merit for clarity.

\subsection{Microwave power}

\begin{figure}[h]
\centering
\includegraphics[width=0.45\textwidth]{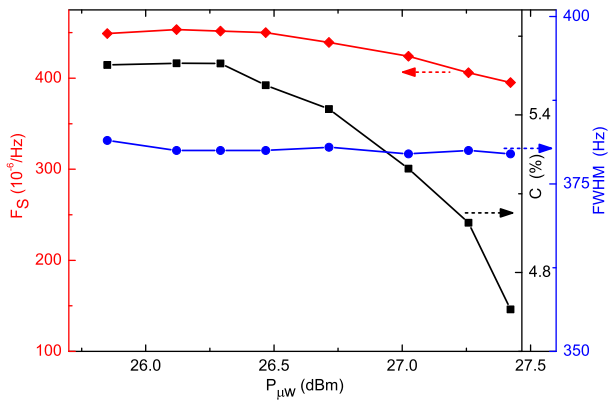}
\caption{(Color online) $F_S$, $C$ and width of the clock transition as function of microwave power $P_{\mu w}$ with 
$T_{cell}=\SI{35.1}{\degreeCelsius}$. All other working parameters is the same as Fig.~10. }
\label{Puw}
\end{figure}

\par $F_S$, $C$ and width versus the microwave power are shown in Fig.~\ref{Puw}.  
The behaviour of  $F_S$ is basically in agreement with the fractional power of first ($\pm1$) sidebands  of Fig.~\ref{sideband}.  
The optimized microwave power is around $\SI{26.12}{\dBm}$.

\subsection{ Detection window $t_w$ and pumping time $t_d$}

\par
\begin{figure}[h]
\centering
\includegraphics[width=0.45\textwidth]{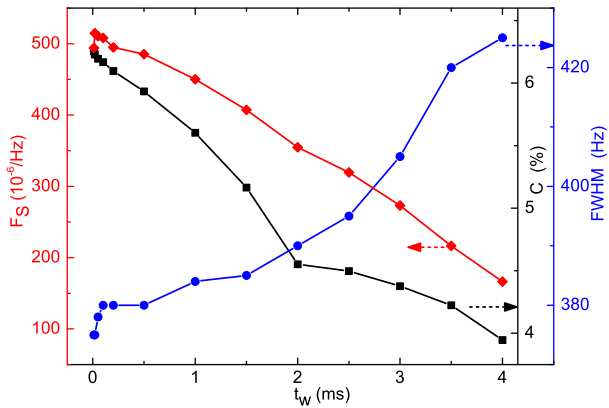}
\caption{(Color online) 
$F_S$, $C$ and width of the clock transition as a function of $t_w$. 
With $t_d= \SI{4}{\milli\second}-t_w$ and all other working parameters is the same as Fig.~10.}
\label{tw}
\end{figure}

 As we can see on Fig.~\ref{tw}, a short detection window $t_w$ would generate a higher contrast signal
and higher figures of merit. 
However, we found that a longer time, e.g., $t_w= \SI{1}{\milli\second}$,
results in a better Allan deviation at one-second averaging time.
It is due to the conflict between the higher signal slope ($\propto F_{S}$) and the increased number
of detected samples which help to reduce the noise, see Eq.(\ref{sigma_yp_i-3}).

\par
\begin{figure}[h]
\centering
\includegraphics[width=0.45\textwidth]{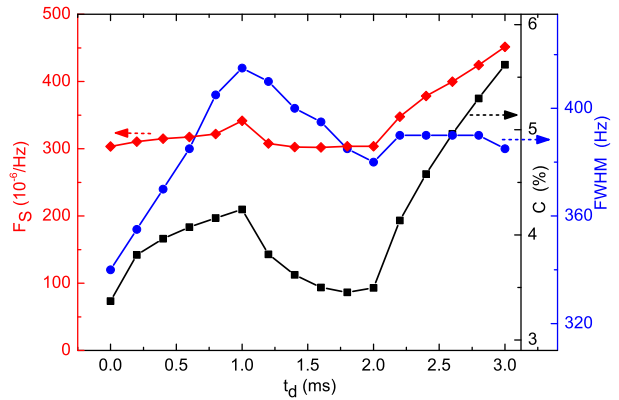}
\caption{(Color online) $F_S$, $C$ and width of the clock transition versus $t_d$. 
All other working parameters is the same as Fig.~10.}
\label{td}
\end{figure}

 The same parameters are plotted versus the pumping time $t_d$ in Fig.~\ref{td}.
The figures of merit and $C$ firstly increase and then decrease at $t_d= \SI{1}{\milli\second}$,
because when the detection window $t_w \geq \SI{1}{\milli\second}$, the signals of two successive polarizations are included.
The dynamic behaviour of the atomic system induces the decrease of the CPT amplitude (see Fig.~10 in \cite{Yun16}).
After $t_d= \SI{2}{\milli\second}$, $F_S$ increases again and reach a maximum.
Thus we can say that, in a certain range, a longer $t_d$ will lead to a greater atomic population pumped into the clock states as
depicted in Fig.~\ref{td}, yielding higher figures of merit. The behaviour of the linewidth versus the pumping time is not explained to date. Nevertheless, note that here the steady-state is not reached and the width behaviour results certainly from a transient effect.

\subsection{Polarization-and-phase modulation frequency $f_m$}

\par
\begin{figure}[h]
\centering
\includegraphics[width=0.45\textwidth]{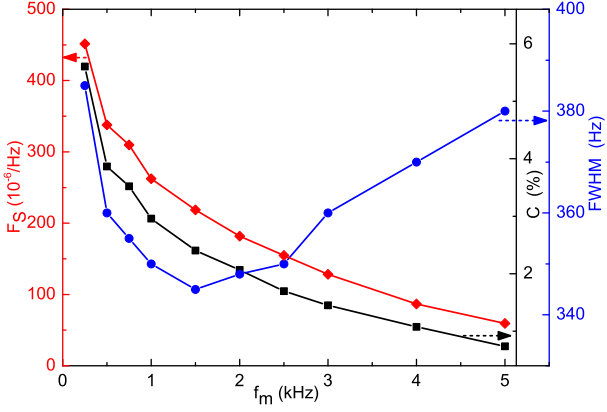}
\caption{(Color online) The $F_S$, $C$ and width of the clock transition as function of $f_m$. 
All other working parameters is the same as Fig.~10.}
\label{fm}
\end{figure}

 Figure \ref{fm} shows $F_S$, $C$, and width versus the polarization (and phase) modulation frequency $f_m$
The maxima of  $F_S$ is reached at low frequency $f_m$.
In one hand, this is an encouraging result to demonstrate the suitability of the LCPR polarization modulator in this experiment. 
In an other hand, the higher $F_M$ rate would be better for a clock operation with lock-in method to modulate
and demodulate the error signal, to avoid the low frequency noises such as $1/f$ noise. 
Therefore, we chose $F_M= \SI{125}{\hertz}$ and $f_m= \SI{250}{\hertz}$.
We have noticed that the behavior of $C$ is not exactly the same than the one observed in our previous work \cite{Yun16,Yun16b} with a fast EOAM, 
where the signal amplitude was maximized at higher frequencies. 
This can be explained by the slower response time of the polarization modulator and the lower laser intensity used.
The linewidth reaches a minimum around $\SI{1.5}{\kilo\hertz}$. This behaviour will be investigated in the future.

\section{Frequency stability}

\subsection{Measured stability}

\begin{figure}[h]
\centering
\includegraphics[width=0.45\textwidth]{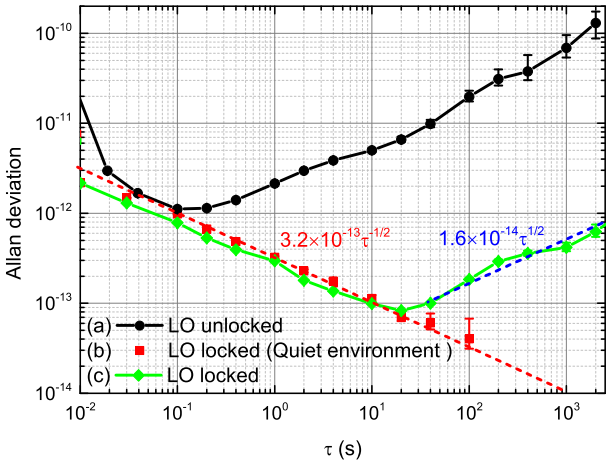}
\caption{(Color online) The LO frequency stability (a) free running, 
(b) and (c) locked on the atomic resonance. (b) best record in quiet environment, (c) typical record.
The slope of the red (blue) dashed fitted line is $3.2 \times 10^{-13} \tau^{-1/2}$ ($1.6 \times 10^{-14} \tau^{1/2}$), respectively. }
\label{Allan}
\end{figure}

The high contrast and narrow line-width CPT signal obtained with the optimized values of the parameters is presented in Fig.~\ref{CPTsignal},
with the related error signal.
The Allan standard deviation of the free-running LO and of the clock frequency, measured against the H maser, are shown in Fig.~\ref{Allan}.
The former is in correct agreement with its measured phase noise. 
In the \SIrange{1}{100}{\hertz} offset frequency region, the phase noise spectrum of the free-running \SI{4.596}{\giga\hertz} LO signal 
is given in \SI{}{\decibel\square\radian/\hertz} by $S_{\varphi}(f) = b_{-3}f^{-3}$ with $b_{-3}$ = $-$47, 
signature of a flicker frequency noise \cite{Francois15}. 
This phase noise yields an expected Allan deviation given by $\sigma_y$(\SI{1}{\second}) $\approx$ $\sqrt{2\ln2\times \frac{4 b_{-3}}{f_c^2}}$ $\approx$ 1.2$\times 10^{-12}$ \cite{Rubiola12}, close to the measured value of $2\times 10^{-12}$ at \SI{1}{\second}.
The measured stability of the CPT clock is $3.2 \times 10^{-13} \tau^{-1/2}$ up to \SI{100}{\second} averaging time for our best record.
This value is close to the best CPT clocks \cite{Hafiz15,Danet14}, demonstrating that a high-performance
CPT clock can be built with the DM-CPT scheme. A typical record for longer averaging times is also
shown in Fig.~\ref{Allan}. For averaging times $\tau$ longer than \SI{20}{\second}, the Allan deviation increases
like $\sqrt{\tau}$, signature of a random walk frequency noise.

\subsection{Short-term stability limitations}

We have investigated the main noise sources that limit the short-term stability. For a first estimation, we consider only white noise sources, and for the sake of simplicity we assume that the different contributions can independently add, so that the total Allan variance can be computed as 
\begin{equation}
\sigma_y^2(\tau) = \Sigma_i \sigma_{y,p_i}^2(\tau)+\sigma_{y,LO}^2(\tau),
\label{sigma_ytotal}
\end{equation}
 with $\sigma_{y,LO}^2(\tau)$ the contribution due to the phase noise of the local oscillator, and $\sigma_{y,p_i}^2(\tau)$ the Allan variance of the clock frequency induced by the fluctuations of the parameter $p_i$. When $p_i$ modifies the clock frequency during the whole interrogation cycle,  $\sigma_{y,p_i}^2(\tau)$ can be written 
 
 \begin{equation}
\sigma_{y,p_i}^2(\tau) = \frac{1}{f_c^2}\left(\sigma_{p_i}^2\right)_{1Hz} \left(\delta f_c/ \delta p_i \right)^2 \frac{1}{\tau},
\label{sigma_yp_i}
\end{equation}
with $\left(\sigma_{p_i}^2\right)_{1Hz}$ the variance of $p_i$ measured in 1 Hz bandwidth at the modulation frequency $F_M$, $\left(\delta f_c/ \delta p_i \right)$ is the clock frequency sensitivity to a fluctuation of $p_i$. 
Here, the detection signal is sampled during a time window $t_w$ with a sampling rate $2F_M=1/T_c$, where $T_c$ is a cycle time. 
In this case Eq.(\ref{sigma_yp_i}) becomes
  \begin{equation}
\sigma_{y,p_i}^2(\tau) = \frac{1}{f_c^2}\left(\sigma_{p_i}^2\right)_{tw} \left(\delta f_c/ \delta p_i \right)^2 \frac{T_c}{\tau},
\label{sigma_yp_i-2}
\end{equation} 
with $\left(\sigma_{p_i}^2\right)_{tw}$ the variance of $p_i$ sampled during $t_w$; $\left(\sigma_{p_i}^2\right)_{tw}\approx S_{p_i}(F_M)/(2t_w)$ with $S_{p_i}(F_M)$ the value of the power spectral density (PSD) of $p_i$ at the Fourier frequency $F_M$ ( assuming a white frequency noise around $F_M$). 
When $p_i$ induces an amplitude fluctuation with a sensitivity $\left(\delta V_{wp}/ \delta p_i \right)$, Eq.(\ref{sigma_yp_i-2}) becomes
 \begin{equation}
\sigma_{y,p_i}^2(\tau) = \frac{1}{f_c^2}S_{p_i}(F_M)\frac{\left(\delta V_{wp}/\delta p_i \right)^2 }{S_\ell^2}  \frac{T_c}{2t_w}\frac{1}{\tau},
\label{sigma_yp_i-3}
\end{equation} 
with $S_\ell$ the slope of the frequency discriminator in V/Hz. 
We review below the contributions of the different sources of noise.

\begin{figure}[h]
\centering
\includegraphics[width=0.45\textwidth]{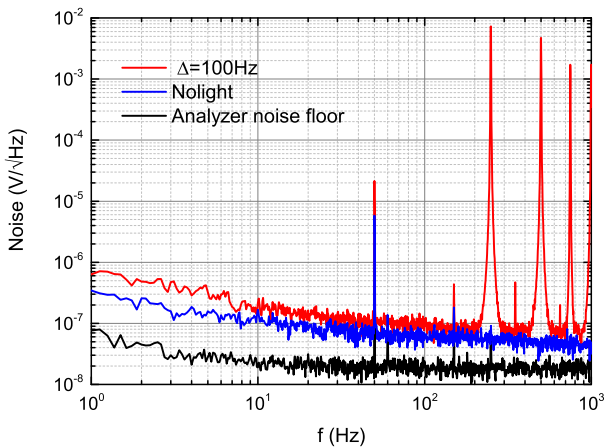}
\caption{(Color online) The RIN after the cell, detector noise and analyser noise floor.
Working parameters is the same as Fig.~10.}
\label{SNR}
\end{figure}

\par Detector noise: the square root of the power spectral density (PSD) $S_{d}$ of the signal fluctuations measured in the dark is shown in 
Fig.~\ref{SNR}. It is
$N_{detector}=64.8$ nV$/\sqrt{\text{Hz}}$ in $\SI{1}{\hertz}$ bandwidth at the Fourier frequency \SI{125}{\hertz}. 
According to Eq.(\ref{sigma_yp_i-3}) the contribution of the detector noise to the Allan deviation at one second is $\num{0.45d-13}$.
 
\par Shot-noise: with the transimpedance gain $G_R=\SI{1.5d4}{\volt\per\ampere}$ and the detector current
$I=V_{wp}/G_R=\SI{32.7}{\micro\ampere}$, Eq.(\ref{sigma_yp_i-3}) becomes \begin{equation}
\sigma_{y,sh}^2(\tau) = \frac{1}{f_c^2}\frac{\left(2eIG_R \right)^2 }{S_\ell^2}  \frac{T_c}{2t_w} \frac{1}{\tau},
\label{sigma_y-sh}
\end{equation} 
with $e$ the electron charge. The contribution to the Allan deviation at one second is \num{0.34d-13}. 

\par Laser FM-AM noise: it is the amplitude noise induced by the laser carrier frequency noise. The slope of the signal $V_{wp}$ with respect to the laser frequency $f_L$ is $S_{FM-AM}=\SI{0.16}{\milli\volt\per\mega\hertz}$ at optical resonance. According to Eq.(\ref{sigma_yp_i-3}) with data of laser-frequency-noise PSD of Fig.~\ref{fig5} at 125 Hz, we get a Allan deviation of \num{0.33d-13} at one second.

\par Laser AM-AM noise: it is the amplitude noise induced by the laser intensity noise. The measured signal sensitivity to the laser power is  $S_{AM}=\SI{3.31}{\milli\volt\per\micro\watt}$ at $f_L=\SI{163}{\micro\watt}$, 
combined with the laser intensity PSD of Fig.~\ref{fig3} it leads to the amplitude noise $S_{AM}\times P_L \times RIN(\SI{125}{\hertz})=30.3$ nV$/\sqrt{\text{Hz}}$, and an Allan deviation of \num{0.21d-13} at one second.

\par LO phase noise: the phase noise of the local oscillator degrades the short-term frequency stability via the intermodulation effect \cite{Audoin91}. 
It can be estimated by:
\begin{equation}
\sigma_{y_{LO}}(1s) \sim \frac{F_M}{f_c}\sqrt{S_{\varphi}(2F_M)}.
\label{PD}
\end{equation}

Our \SI{4.596}{\giga\hertz} microwave source is based on \cite{Francois15} which shows an ultra-low phase noise
$S_{\varphi}(2F_M)=\SI{-116}{\decibel\square\radian\per\hertz} $
at $2F_M=\SI{250}{\hertz}$ Fourier frequency. This yields a contribution to the Allan deviation of \num{0.43d-13} at one second.

\par Microwave power noise: fluctuations of microwave power lead to a laser intensity noise, which is already taken into account in the RIN measurement. We show in the next section that they also lead to a frequency shift (see Fig.~\ref{f0-Puw}). 
The Allan deviation of the microwave power at $\SI{1}{\second}$ is $\SI{2.7d-4}{\dBm}$, see inset of  Fig.~\ref{f0-Puw}. With a measured slope of $7.7$ Hz/dBm, we get a fractional-frequency Allan deviation of \num{2.26d-13}, which is the largest contribution to the stability at \SI{1}{\second}. Note that in our set-up the microwave power is not stabilized. \\

The other noise sources considered have much lower contributions, they are the laser frequency-shift effect,
\textit{i.e.} AM-FM and FM-FM contributions, the cell temperature and the magnetic field. Table \ref{1s} resumes the short-term stability noise budget.

\begin{table}[h]
\caption{\label{tab:table1} Noise contributions to the stability at \SI{1}{\second}.}
\begin{ruledtabular}
 \begin{tabular}{ccc}
Noise source              & noise level                                     & $\sigma_y(1s)\times\num{d13}$ \\
\hline
Detector noise            &  $\SI{64.8}{\nano\volt}/\sqrt{\text{Hz}}$           & \num{0.45} \\
Shot noise                &  $\SI{48.8}{\nano\volt}/\sqrt{\text{Hz}}$           & \num{0.34} \\
Laser FM-AM               &  $\SI{48.0}{\nano\volt}/\sqrt{\text{Hz}}$           & \num{0.33}  \\
Laser AM-AM               &  $\SI{30.3}{\nano\volt}/\sqrt{\text{Hz}}$           & \num{0.21}   \\
LO phase noise            &  $\SI{-116}{\decibel\square\radian/\hertz}$         & \num{0.43}    \\
$P_{\mu w}$               &  $\SI{2.7d-4}{\dBm}@\SI{1}{\second}$                & \num{2.26}    \\
Laser AM-FM               &  $\SI{0.6}{\nano\watt}@\SI{1}{\second}$             & \num{9.7d-3}    \\
Laser FM-FM               &  $\sim\SI{100}{\hertz}@\SI{1}{\second}$             & \num{2.9d-3}    \\
$T_{cell}$                &  $\SI{6.3d-5}{\kelvin}@\SI{1}{\second}$             & \num{3.2d-2}    \\
$B_0$                     &  $\SI{4.3}{\pico\tesla}@\SI{1}{\second}$            & \num{1.4d-3}    \\
\hline
Total                     &                                                     & \num{2.4}     \\
\end{tabular}
\end{ruledtabular}
\label{1s}
\end{table}

\par The laser intensity noise after interacting with the atomic vapor is depicted in Fig.~\ref{SNR}.
It encloses the different contributions to the amplitude noise, \textit{i.e.} detector noise, shot-noise,
FM-AM and AM-AM noises. The noise spectral density is $\SI{100}{\nano\volt}/\sqrt{\text{Hz}}$ at the Fourier frequency \SI{125}{\hertz},
which leads to an Allan deviation of \num{0.7d-13} at \SI{1}{\second}.
This value is equal  to the quadratic sum of the individual contributions.
The quadratic sum of all noise contribution leads to an Allan deviation at one second of
\num{2.4d-13}, while the measured stability is $3.2\times 10^{-13}\tau^{-1/2}$ (Fig.~\ref{Allan}).
The discrepancy could be explained by correlations between different noises, which are not all independent.
The dominant contribution is the clock frequency shift induced by the microwave power fluctuations.
This term could be reduced by microwave power stabilization or a well chosen laser
power (see Fig.~\ref{Puw}), but to the detriment of the signal amplitude.

\section{Frequency shifts and mid-term stability}

We have investigated the clock-frequency shift with respect to the variation of various parameters. 
For each frequency measurement, the LO frequency is locked on the CPT resonance. With a \SI{100}{\second} averaging time, the mean frequency is measured with typical error bar less than \num{d-12}, \textit{i.e.}, 0.01 Hz relative to the Cs frequency $f_{Cs}= \SI{9.192}{\giga\hertz}$.

\subsection{$f_0$ vs $T_{cell}$}

\begin{figure}[h]
\centering
\includegraphics[width=0.45\textwidth]{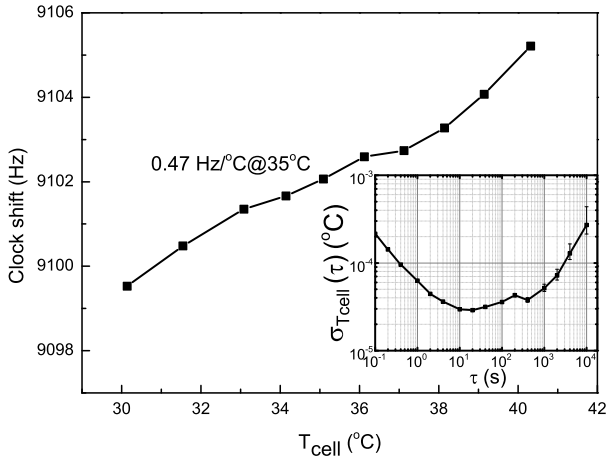}
\caption{(Color online)	Clock frequency as a function of the cell temperature $T_{cell}$. 
The inset shows the Allan deviation of $T_{cell}$. All other working parameters is the same as Fig.~10.}
\label{f0-Tcell}
\end{figure}

The resonance frequency of the microwave resonance is shifted by collisions between Cs atoms and buffer-gas atoms \cite{Vanier89}.
This collisional shift is temperature dependent but can be reduced by using a well-chosen mixture of gas. 
Here we use a N$_2$-Ar mixture with 37$\%$ of Ar,
optimized for cancelling the temperature coefficient at \SI{31.5}{\degreeCelsius} and which should allow to reduce the sensitivity of the cock frequency at the level of \SI{-0.16}{\hertz\per\kelvin} at $\SI{35}{\degreeCelsius}$ \cite{Kozlova11}.
The record of the frequency shift versus $T_{cell}$ is presented in Fig.~\ref{f0-Tcell}. 
The temperature sensitivity is measured to be
$\SI{0.47}{\hertz\per\kelvin}$ at $T_{cell}= \SI{35}{\degreeCelsius}$. 
This value is in disagreement with the expected value. Nevertheless, it is important to note that the latter is valid at null laser power and that the laser power shift is also temperature-dependent \cite{Kozlova14}.
In this experimental test, we measure the result of the collisional shift and of the laser power shift together. 
The Allan deviation of $T_{cell}$ is shown in the inset of Fig.~\ref{f0-Tcell}. 
With a typical temperature fluctuation of $ \num{5.2d-5}\SI{}{\kelvin}$ at \SI{1000}{\second}, 
the contribution of cell temperature variations to the clock fractional frequency stability is about  \num{2.7d-15} at \SI{1000}{\second}.

\subsection{$f_0$ vs $P_L$}

\par
\begin{figure}[h]
\centering
\includegraphics[width=0.45\textwidth]{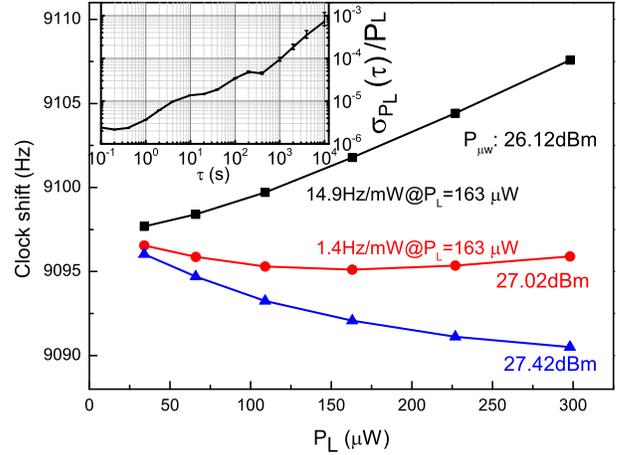}
\caption{(Color online)	Clock frequency as a function of laser power $P_L$ for different values of  $P_{\mu w}$. 
Inset: fractional Allan deviation of the laser power. All other working parameters is the same as Fig.~10. }
\label{f0-PL}
\end{figure}

 The clock frequency shift versus $P_L$ is presented in Fig.~\ref{f0-PL}. The coefficient of the light power shift is
$\SI{14.9}{\hertz\per\milli\watt}$ at $P_L= \SI{163}{\micro\watt}$ and $P_{\mu w}= \SI{26.12}{\dBm}$. This shift is difficult to foresee theoretically because it results of the combination of light shifts (AC Stark shift) induced by all sidebands of the optical spectrum, but also of overlapping and broadening of neighboring lines. The inset of  Fig.~\ref{f0-PL} shows the typical fractional fluctuations the laser power versus the integration time. They are measured to be $\num{1d-4}$ at \num{1000} s, impacting on the clock fractional frequency stability at the level of \num{2.6d-14} at \SI{1000}{\second}. Since the power distribution in the sidebands vary with the microwave power, the laser power shift is also sensitive to the microwave power feeding the EOPM. This is clearly shown in Fig.~\ref{f0-PL}. As previously observed in CPT-based clocks \cite{Zhu00,Levi00} 
and double-resonance Rb clocks \cite{Affolderbach05}, 
it is important to note that the light-power shift coefficient can be decreased and even cancelled at specific values of $P_{\mu w}$. 
Consequently, it should be possible to improve the long-term frequency stability by tuning finely the microwave power value 
\cite{Shah:APL:2006, Happer:APL:2009}, at the expense of a slight degradation of the short-term frequency stability.

\subsection{$f_0$ vs $P_{\mu W}$}

\begin{figure}[h]
\centering
\includegraphics[width=0.45\textwidth]{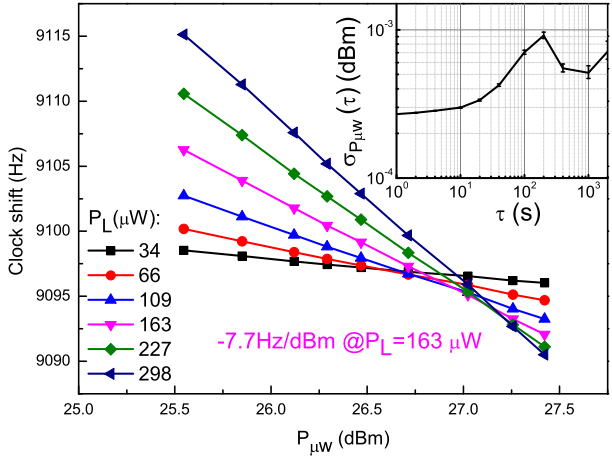}
\caption{(Color online)	Clock frequency as a function of the microwave power $P_{\mu w}$. 
Inset: Allan deviation of the microwave power in dBm, log scale. All other working parameters is the same as Fig.~10.}
\label{f0-Puw}
\end{figure}

The frequency shift versus the microwave power at fixed laser power is shown on Fig.~\ref{f0-Puw}.
At constant optical power, only the power distribution among the different sidebands changes. 
This is a different laser power effect.
The shift scales as the microwave power in the investigated range, 
with a sensitivity of  $\SI{7.7}{\hertz\per\dBm}$ at $P_L = \SI{163}{\micro\watt}$.
In this range, the power ratio of both first ($\pm 1$) sidebands changes by about $10\%$. 
The inset of Fig.~\ref{f0-Puw} shows the Allan deviation of the microwave power in dBm. The typical microwave power standard deviation of 
$\SI{5d-4}{\dBm}$ at \SI{1000}{\second} yields a  fractional frequency stability of about \num{4.2d-13} at \SI{1000}{\second}.

\subsection{$f_0$ vs $\Delta_L$}

\begin{figure}[h]
\centering
\includegraphics[width=0.45\textwidth]{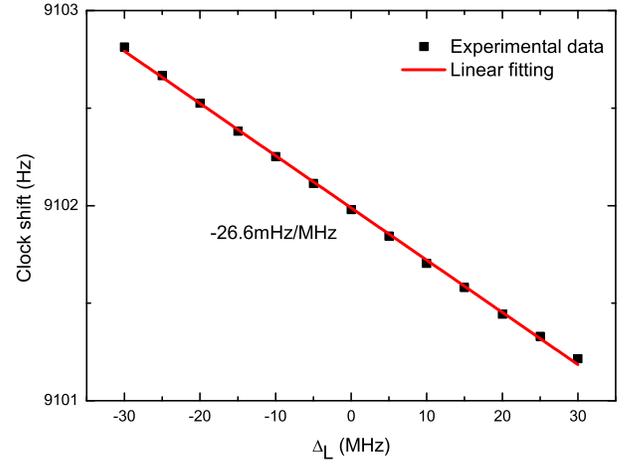}
\caption{(Color online)	Clock frequency as a function of laser frequency detuning $\Delta_L$.
Experimental data are fitted by a linear function. All other working parameters is the same as Fig.~10.}
\label{f0-fL}
\end{figure}

The frequency of the laser beam incident on the clock cell can be tuned by setting the driving frequency of AOM2 in the laser frequency stabilization setup. For each AOM driving frequency, the laser carrier frequency is stabilized onto the reference cell, 
the clock frequency is locked onto the CPT resonance and measured against the hydrogen maser. 
The observed shift results from a combination of AC Stark shift, CPT resonance distortion and effect of neighboring lines.
The recorded frequency shift is shown on Fig.~\ref{f0-fL} versus the laser detuning.
The shift is well-fitted by a linear function with a slope of \SI{-26.6}{\milli\hertz\per\mega\hertz}. 
Having no second similar DFB diode laser set-up, we did not measure the laser frequency stability. 
Taking into account that we have the same diode laser and laser frequency stabilization setup than the one reported in \cite{Hafiz16}, 
we expect similar performances, \textit{i.e.}, a standard deviation of $\sim$\SI{5}{\kilo\hertz} at \num{1000} second of averaging time. This yields a clock-fractional-frequency stability of about \num{1.4d-14} at \SI{1000}{\second}.

\subsection{$f_0$ vs $f_m$}

\begin{figure}[h]
\centering
\includegraphics[width=0.45\textwidth]{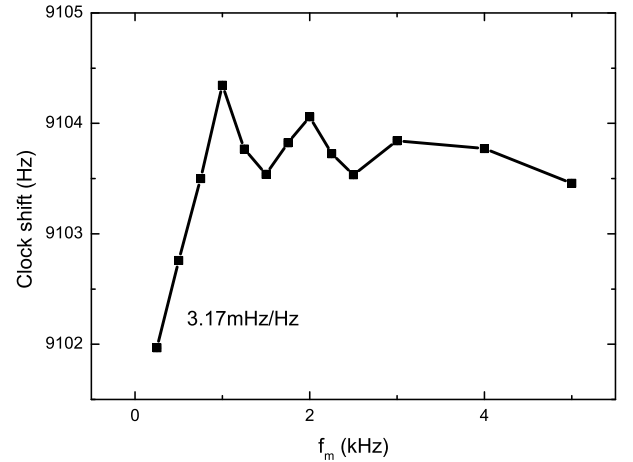}
\caption{(Color online)	Clock frequency as function of polarization(phase) modulation frequency $f_m$. 
All other working parameters is the same as Fig.~10.}
\label{f0-fm}
\end{figure}

For the sake of completeness, we have measured the clock shift versus polarization (and phase) modulation frequency $f_m$. 
Results are reported in Fig.~\ref{f0-fm}. The other parameters are fixed.  
The shift coefficient is $\SI{3.17}{\milli\hertz\per\hertz}$ at $f_m=\SI{250}{\hertz}$.
As $f_m$ is synchronized to the LO, which exhibits in the worst case (unlocked) a frequency stability at the level of
\num{7d-11} at  \SI{1000}{\second}  (see Fig.~\ref{Allan}), the effect of the polarization and phase modulation frequency on the clock shift is negligible, \textit{i.e.} \num{6.0d-21}  at  \SI{1000}{\second} second.

\subsection{$f_0$ vs $B_0$}

\begin{figure}[h]
\centering
\includegraphics[width=0.45\textwidth]{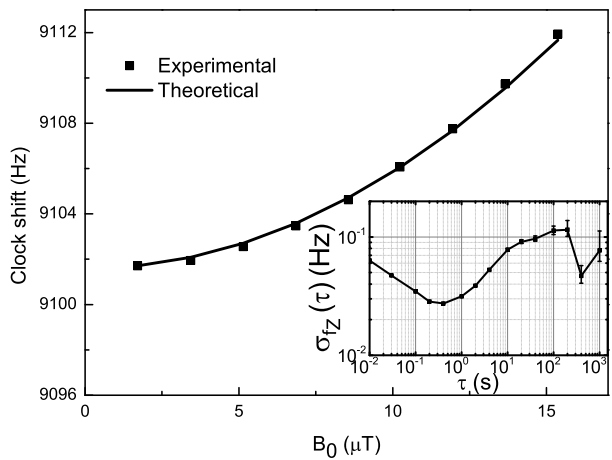}
\caption{(Color online)	Clock frequency as function of magnetic field $B_0$. Inset: Allan deviation of the Zeeman frequency. 
All other working parameters is the same as Fig.~10.}
\label{f0-B0}
\end{figure}

The clock frequency shift versus the magnetic field strength is shown on Fig.~\ref{f0-B0}. 
The experimental shift is in good agreement with the theoretical prediction of the quadratic Zeeman shift \cite{Vanier89}, $f_0= \num{4.27d-2} B_0^2$, with $B_0$ in \SI{}{\micro\tesla}.
In order to measure the time evolution of the magnetic field, we locked the LO frequency to the magnetic-field sensitive Zeeman CPT transition
$\ket{F=3,m_F=-1}\rightarrow\ket{F=4,m_F=-1}$. 
The latter exhibits a sensitivity \SI{7.01}{\kilo\hertz\per\micro\tesla}.
The Allan standard deviation of the Zeeman frequency ($f_Z$) is shown in the inset of Fig.~\ref{f0-B0}. 
The measured frequency deviation is about \SI{0.1}{\hertz} at \SI{1000}{\second}, 
corresponding to a magnetic field deviation of \SI{14}{\pico\tesla} at $\tau=\SI{1000}{\second}$. 
For a mean magnetic field of \SI{3.43}{\micro\tesla}, this yields  a fractional frequency stability of
\num{0.4d-15} at \SI{1000}{\second}.

\subsection{Mid-term stability}

With the shift coefficients and the Allan standard deviation of the involved parameters, we can estimate the various contributions to the mid-term clock frequency stability. They are listed in Table \ref{1000s} for a \SI{1000}{\second} averaging time. 
Their quadratic sum leads to a frequency stability of \num{4.2d-13} at $\tau=\SI{1000}{\second}$, 
in very good agreement with the measured stability \num{4.21d-13} (see Fig.~\ref{Allan}).
Again, the main contribution to the instability comes from the microwave power fluctuations, before the laser power and frequency fluctuations. Thus in the future, it is necessary to stabilize the microwave power to improve both the short-and-mid-term frequency stability.

\begin{table}[h]
\caption{\label{tab:table2}Noise contributions to the stability at \SI{1000}{\second}.}
\begin{ruledtabular}
\begin{tabular}{ccc}
Parameter              & coefficient                                          & $\sigma_y(1000s)\times\num{d13}$ \\
\hline
$T_{cell}$                  &  $\SI{0.47}{\hertz\per\kelvin}$                 &   \num{2.7d-2}  \\
$P_L$                       &  $\SI{14.9}{\hertz\per\milli\watt}$             &   \num{0.26}         \\
$P_{\mu w}$                 &  $\SI{-7.7}{\hertz\per\dBm}$                    &   \num{4.2}             \\
$\Delta_L$                  &  $\SI{-26.6}{\milli\hertz\per\mega\hertz}$  &   \num{0.14}   \\
$B_0$                       &  $\SI{0.29}{\hertz\per\micro\tesla}$            &   \num{0.4d-2}       \\
\hline
Total                       &                                                 &   \num{4.2}   \\
\end{tabular}
\end{ruledtabular}
\label{1000s}
\end{table}

\section{Conclusions}
\par We have implemented a compact vapor cell atomic clock based on the DM CPT technique. 
A detailed characterization of the CPT resonance versus several experimental parameters was performed. 
A clock frequency stability of $3.2 \times 10^{-13} \tau^{-1/2}$ up to \SI{100}{\second} averaging time was demonstrated. 
For longer averaging times, the Allan deviation scales as $\sqrt{\tau}$, signature of a random walk frequency noise. 
It has been highlighted that the main limitation to the clock short and mid-term frequency stability is the fluctuations of the microwave power feeding the EOPM. Improvements could be achieved by implementing a microwave power stabilization. 
Another or complementary solution could be to choose a finely tuned laser power value minimizing the microwave power sensitivity. 
This adjustment could be at the expense of the signal reduction and a trade-off has to be found. 
Nevertheless, the recorded short-term stability is already at the level of best CPT clocks \cite{Danet14, Hafiz15} and close to state-of-the art Rb vapor cell frequency standards. These preliminary results show the possibility to a high-performance and compact CPT clock based on the DM-CPT technique.

\section*{Acknowledgements}
\par We thank Moustafa Abdel Hafiz (FEMTO-ST), David Holleville and Luca Lorini (LNE-SYRTE) for helpful discussions. 
We are also pleased to acknowledge Charles Philippe and Ouali Acef for supplying the thermal insulation material,
Michel Abgrall for instrument Symmetricom 5125A lending,
David Horville for laboratory  arrangement,
Jos\'e Pinto Fernandes, Michel Lours for electronic assistance,
Pierre Bonnay and Annie G\'erard for manufacturing Cs cells.

P. Y. is supported by the Facilities for Innovation, Research, Services, 
Training in Time \& Frequency (LabeX FIRST-TF). This work is supported in part
by ANR and DGA(ISIMAC project ANR-11-ASTR-0004). This work has been funded by
the EMRP program (IND55 Mclocks). The EMRP is jointly funded by the EMRP participating
countries within EURAMET and the European Union.


\end{document}